\documentclass[dvipdfmx,12pt]{article}
\usepackage{amsmath,amsfonts,amssymb,booktabs,multirow,cite}
\usepackage{array}
\usepackage[dvipdfmx]{graphicx}
\usepackage{xcolor}

\makeatletter

\def\affiliation#1{\def\@affiliation{#1}}

\def\@maketitle{
\begin{center}
{\Large \bf \@title \par}
\vspace{5mm}
{\large \@author \par}
{\normalsize \it \@affiliation \par}
\end{center}
}

\@addtoreset{equation}{section}

\title{A note on no-hair properties of static black holes in four and higher dimensional spacetimes with cosmological constant}

\medskip 

\author{Akihiro Ishibashi${}^{1,2}$, Satoshi Matsumoto${}^1$, and Yuichiro Yoneo${}^1$} 

{}\bigskip 

\affiliation{${}^{1}$Department of Physics and ${}^{2}$Research Institute for Science and Technology, \\ 
Kindai University, Higashi-Osaka, Osaka 577-8502, Japan} 

\medskip 

\date{}

\begin{document}

\maketitle

\abstract{
We study no-hair properties of static black holes in four and higher dimensional spacetimes with a cosmological constant. For the vanishing cosmological constant case, we show a no-hair theorem and also a no-short-hair theorem under certain conditions for the energy-momentum of matter fields. For the positive cosmological constant case, we discuss conditions for hairy static black holes to exist in terms of the energy density of matter fields evaluated at the black hole horizon and the cosmological horizon. For the negative cosmological constant case, we study conditions for hairy black holes by presenting a no-hair theorem in which the asymptotic structure is assumed to be determined by the true cosmological constant. 
}

\section{Introduction} 

It is well known that a stationary black hole in general relativity has the uniqueness or no-hair property: All asymptotically flat, stationary black holes in $4$-dimensional Einstein-Maxwell theory are uniquely characterized by their mass, angular momentum, and electric charge\cite{Israel:1967wq,Israel:1967za,Carter:1971zc,Robinson:1974nf,Robinson:1975bv,Mazur:1982db} (see also a review~\cite{Chrusciel:2012jk}). 
No-hair properties have so far been studied in a number of models with various different matter fields, by either proving no-hair theorems~\cite{Bekenstein:1971hc,Pena:1997cy,Nunez:1996xv,Lin:2020asf,Matsumoto:2022zvj} or explicitly (analytically or numerically) constructing hairy black hole solutions~\cite{Bizon:1990sr,Garfinkle:1990qj,Bizon:1992gb,Lavrelashvili:1992ia,Greene:1992fw,Nucamendi:1995ex,Herdeiro:2018wub,Fernandes:2019rez,Santos:2020pmh}. 
Here, by ``hair" we mean some non-trivial structure of matter fields outside the black hole event horizon which are not specified in terms of the global conserved charges measured at asymptotic infinity. As a more modest alternative to the no-hair theorems, it has been conjectured---in fact, shown for certain cases~\cite{Nunez:1996xv}---that if a black hole is allowed to possess hair, the hairy region must extend beyond $3/2$ the horizon radius. This is known as the no-short hair conjecture. Most of such studies have focused mainly on the case of $4$-dimensional, asymptotically flat spacetimes.  

There have recently been plenty of astrophysical and cosmological evidence to support unknown dark sector in our universe. The existence of dark matter and dark energy (or a positive cosmological constant) modifies, in principle, the spacetime structure near the event horizon, as well as asymptotic region, which allows for a cosmological horizon. It is of considerable importance to study no-hair properties of stationary black hole spacetimes with positive cosmological constant (see, e.g., \cite{MaAYu15,Borghini:2019msu} for no-hair theorems and also e.g.,~\cite{Cai:1997ij,An:2023bpb} for examples of hairy black holes in asymptotically de Sitter spacetimes).

A black hole spacetime also plays an important role as a theoretical laboratory for various theories of gravity, such as string theories, in which the notion of higher dimensional spacetime or extra-dimension is essential. It is known that in higher dimensions, the standard notion of black hole no-hair properties no longer holds (see, e.g., reviews \cite{Emparan:2008eg,Hollands:2012xy,Ida:2011jm}). Another interesting theoretical possibilities are black hole spacetimes with negative cosmological constant, which have attracted much attention in the context of AdS/CFT correspondence or more broadly holographic principle~\cite{Maldacena:1997re,Gubser:1998bc,Witten:1998qj}. It is therefore of great interest in considering no-hair properties of black holes in higher dimensional spacetimes with cosmological constant either positive or negative (see e.g., \cite{Cai:1997ij,An:2023bpb,Torii:2001pg,Winstanley:1998sn,Winstanley:2002jt,Brihaye:2019gla,Sudarsky:2002mk} for the study of no-hair properties with a non-vanishing cosmological constant in $4$-dimensions).  

The main purpose of this paper is to analyze no-hair properties of static black holes in four and higher dimensional Einstein gravity with cosmological constant. For asymptotically locally flat case, we show a no-hair theorem and also a no-short-hair theorem, generalizing to higher-dimensions 
Pena-Sudarsky's no-hair theorem~\cite{Pena:1997cy} and Nunez-Quevedo-Sudarsky's no-short hair theorem~\cite{Nunez:1996xv}. For positive cosmological constant case, we discuss conditions for hairy static black holes to exist in terms of the energy density of matter fields evaluated at the black hole horizon and the cosmological horizon. For asymptotically AdS case, we study conditions for hairy black holes by generalizing to higher-dimensions Sudarsky-Gonzalez' theorem~\cite{Sudarsky:2002mk}. 

The outline of this paper is as follows. In the next section, we summarize our geometric ansatz and basic requirements for matter fields. In Sec.~\ref{sec:3}, we present our main theorems and their proofs. Section~\ref{sec:4} is devoted to summary and discussions. In Appendix~A, 
we examine the energy conditions of a one-parameter family of asymptotically flat, static, spherically symmetric metrics. 
In Appendix~B, 
we consider the relation between the no-short hair condition and the photon-sphere.

\section{Static black hole metrics and matter fields}
\label{sec:2}

We consider the $(2+n)$-dimensional metric of the following warped product form 
\begin{equation}
\label{metric_n}
ds^2 =g_{\mu \nu}dx^\mu dx^\nu = - \mu(r) e^{-2 \delta(r)} dt^2 + \frac{1}{\mu(r)} dr^2 + r^2 \gamma_{ij}(z) dz^i dz^j \,,
\end{equation}
where $\mu$ and $\delta$ are functions of $r$ only and $\gamma_{ij}$ denotes the metric of the $n$-dimensional Einstein manifold ${\cal K}_n$ with normalized sectional curvature $k= 1, 0, -1$.  
Here and hereafter the greek indices $\mu, \nu,...$ denote, as usual, tensor on the $(2+n)$-dimensional spacetime, while the latin indices $i,j,...$ tensors on ${\cal K}_n$.

Note that the $n$-dimensional Einstein manifold ${\cal K}_n$ is defined as a solution to 
\begin{equation}
\label{metric_Einstein}
 \hat{R}_{ij} = k (n-1) \gamma_{ij} \,, 
\end{equation}
where ${\hat R}_{ij}$ denotes the Ricci curvature with respect to the metric $\gamma_{ij}$. It is known that for $n=2,3$, ${\cal K}_n$ is a maximally symmetric space with the sectional curvature $k$, while for $n \geq 4$, ${\cal K}_n$ can admit a much larger variety of topologies and metrics. For example, when $n \geq 4$, ${\cal K}_n$ with positive curvature $k=1$ can be a connected sum of an $n$-dimensional round sphere $S^n$, a product manifold of the type $S^m \times S^{n-m} (n > m \geq 2$), or some quotient space there of, thus admitting various different topologies. Furthermore, even for the same spherical topology $S^n$, ${\cal K}_n$ with $n \geq 5$ can admit different metrics, other than the round-sphere metric (see, e.g.,~\cite{Gibbons:2002av}). Thus, our metric ansatz of the factorized warped product form above with two unknown functions $(\mu, \delta)$ can describe a fairly generic class of static black holes in higher dimensions\footnote{In $4$-dimension ($n=2$), the most general cohomogeneity-one, static black holes coupled with uncharged scalar fields are given by \cite{Anabalon:2012ta}. Our metric (\ref{metric_n}) above takes essentially 
the same form as the one given in \cite{Anabalon:2012ta}, up to radial coordinate transformation.}{}\footnote{Further generalization within the same context could be to consider the case in which the metric $\gamma_{ij}$ is a function not only of the angle coordinates $z^i$ but also of the radial coordinate $r$, so that the warp factor is no longer factorizable. However, under such a non-factorizable metric ansatz, the equations of motion become highly involved. In this respect, it is interesting to note that recently an infinite number of static, maximally symmetric non-warped product type metrics have been numerically found in higher dimensional asymptotically AdS spacetimes~\cite{Horowitz:2022hlz}. }

We consider the Einstein equations with a cosmological constant $\Lambda$ so that the above metric is a solution to  
\begin{align}
\label{einstein_n}
\begin{split}
G^\mu_\nu &= 8\pi {T^{(\rm tot)}}^\mu_\nu := 8\pi {T}^\mu_\nu - \Lambda \delta^\mu_\nu \,, 
\end{split}
\end{align}
where $T^\mu_\nu$ denotes the energy-momentum tensor (EMT) for matter fields. 
Note that in general one cannot uniquely separate the EMT for matter fields and cosmological constant term. In what follows we assume that any 
constant part of matter EMT proportional to the metric is already absorbed in $\Lambda$. 

The above metric \eqref{metric_n} can describe a fairly large class of static black holes, including generalized Schwarzschild-(Anti-)de Sitter black holes. In the following we assume that the metric function $\mu(r)$ admits a single or multiple single-root(s), which corresponds to a non-degenerate Killing horizon(s). When $\Lambda \leq 0$, we further assume that for the largest root $r_b$ of $\mu(r)$, $\delta(r_b)< \infty$ and $\mu(r)>0$ for $r>r_b$ so that $r_b$ corresponds to a black hole event horizon. When $\Lambda >0$, we assume that there exist at least two roots $r_b, r_c(>r_b)$ for which $\delta(r_b)$ and $\delta(r_c)$ are both finite and $\mu(r)>0$  for $r_b<r<r_c$ so that $r_b$ and $r_c$ correspond to the black hole horizon and cosmological horizon, respectively, in an asymptotically de Sitter spacetime. In the following, for either asymptotically flat, AdS, or de Sitter case, 
we refer to the static region outside the black hole where $r_b<r(<r_c)$ and $\mu(r)>0$ 
as the domain of outer communication (DOC)\footnote{DOC is usually defined by $J^-({\cal I}^+)\cap J^+({\cal I}^-)$. We are, however, interested mainly in the static region bounded by the black hole horizon and cosmological horizon when $\Lambda>0$. 
In this paper, by DOC we mean the region $J^-(\gamma)\cap J^+(\gamma)$, where $\gamma$ is any complete timelike orbit of the Killing vector field $\partial/\partial t$. Note that this definition excludes the region outside the cosmological horizon when our spacetime is asymptotically locally de Sitter.  
}. We often denote the horizon radii $r_b$ and $r_c$ collectively by $r_H$.  

It is convenient to describe the metric function $\mu(r)$ in terms of the generalized Misner-Sharp mass function, $m(r)$, as~\cite{Maeda:2006pm},   
\begin{equation}
\label{lapse_n}
\mu (r) = k - \frac{2 m(r)}{r^{n-1}} - \lambda r^2 \,, 
\end{equation} 
with $\lambda := 2\Lambda/n(n+1)$. 
We assume that at large distances ($r \rightarrow \infty$), the mass function behaves as 
\begin{equation}
\label{condi:m:asympt}
m(r) = m_0 + O(1/r^\alpha) \,,
\end{equation} 
with some constants, $m_0 \geq 0 $ and $\alpha>0$, so that the asymptotic structure of \eqref{metric_n} near infinity is determined solely by $\lambda$, $m_0$, and ${\cal K}_n$. 
Then, for instance, when $\lambda=0$ and ${\cal K}_n$ is the round sphere, the ADM mass is given by $M_{\rm ADM}=nm_0{A}_n/8 \pi$, where ${A}_n=2\pi^{(n+1)/2}/\Gamma[(n+1)/2]$ is the area of a unit $n$-sphere. 
We also require that the generalized Misner-Sharp mass is positive in DOC: 
\begin{equation}
m(r) > 0 \quad {\rm for} \quad r_b \leq r ( \leq r_c) \,.
\label{condi:m}
\end{equation}
 
Note that this positivity requirement is not an obvious one. For instance, for the vacuum case, the Misner-Sharp mass is given by the following constant, 
\begin{equation}
 m(r) = \dfrac{ \left[ k + (-\lambda)r_b^2 \right] r_b^{n-1}}{2} \,. 
\end{equation}
We can find that when $\lambda <0$ and $k=-1$ (i.e., AdS black hole with hyperbolic horizon), depending upon the value of $\lambda$ and the horizon radius $r_b$, it is possible for the Misner-Sharp mass $m(r)$ to take a negative value $m(r_b)=[(-\lambda)r_b^2-1] r_b^{n-1}/2<0$ in DOC $r>r_b$, where $\mu(r)>0$. Now, consider a matter field that satisfies the weak energy condition (WEC) in DOC, but vanishes in a neighborhood of the horizon $r_b$. Then, since under the WEC, $m(r)$ becomes a non-decreasing function, as can be seen below eq.~(\ref{m:prime}), we can arrange the matter field configuration in such a way that starting from the negative value $m(r_b)$ at the horizon, $m(r)$ is increasing toward large distances, becomes positive at some point in DOC, and eventually behaves as eq.~(\ref{condi:m:asympt}) with some positive $m_0$ in the asymptotic region. Our requirement~(\ref{condi:m}) excludes such cases. 

Note also that there have been several different types of quasi-local masses proposed so far, mainly in $4$-dimensions~\cite{Szabados:2009eka}, of which positivity has also been studied for certain cases, e.g., \cite{Liu:2003bx,Liu:2004dc}.

\medskip 

We note that from the symmetry structure of the metric ansatz \eqref{metric_n}, the EMT for matter fields takes the form 
\begin{equation}
 T^t_i = T^r_i= 0 \,, \quad T^i_j = P \delta^i_j \,, 
\end{equation}
with $P$ being some function of $r$, corresponding to a tangential pressure along ${\cal K}_n$. We also introduce the energy density $\rho$ and 
the radial pressure $P_r$ by 
\begin{equation}
\rho:= -T^t_t \,, \quad P_r := T^r_r \,.
\end{equation}
In order for the asymptotic structure to be determined by $\Lambda$, we require that 
\begin{equation}
\label{condi:asympt}
\rho, P_r \,, P \sim O(1/r^{n + 2}) \rightarrow 0 \quad \mbox{as $r \rightarrow \infty$} \,. 
\end{equation} 
Note that each mixed component $T^\mu_\nu$ of the EMT for static Maxwell fields in $(2+n)$-dimensions decays faster than \eqref{condi:asympt}, except for the $4$-dimension ($n=2$) case.  

With the assumptions~\eqref{lapse_n}, \eqref{condi:m:asympt}, \eqref{condi:asympt} above, the spacetime~\eqref{metric_n} approaches asymptotically locally either flat, de Sitter, or AdS, depending on $\lambda$, and one may anticipate that whenever the matter EMT identically vanishes, the static black hole under consideration coincides with the generalized Schwarzschild-(A)dS black hole specified by $\lambda$, $m_0$, ${\cal K}_n$. This is indeed the case for $\lambda=0$ in general dimensions as shown in~\cite{Gibbons:2002av}. 
Then, we can say that a static black hole has ``hair" when it allows to have a non-trivial matter and metric configurations outside the black hole event horizon, being different from the generalized Schwarzschild-(A)dS metric, but still asymptotically approaches the corresponding generalized Schwarzschild-(A)dS metric. 

\bigskip 

We impose, as a physically reasonable requirement for EMT, the weak energy condition (WEC) on the matter fields: 
\begin{equation}
\label{wec_n}
\rho \geq 0 ,\quad \rho + P_r \geq 0 , \ \rho + P \geq 0 \,. 
\end{equation}
From the Einstein equations \eqref{einstein_n}, we obtain 
\begin{align}
\label{divmu_n}
\mu^\prime & = - \frac{16\pi r}{n} \rho - (n+1) \lambda r + \frac{\left( n - 1 \right) \left( k - \mu \right)}{r} \,, \\
\label{divdelta_n}
\delta^\prime & = - \frac{8\pi r}{n \mu} \left( \rho + P_r \right) \,,
\end{align}
where the {\it prime} denotes the derivative with respect to $r$. Note that the corresponding formulas for the Einstein-Gauss-Bonnet gravity with $k=1$ have been previously derived in~\cite{Ghosh:2023kge}.
From \eqref{lapse_n}, \eqref{divmu_n} and WEC, we can easily find
\begin{equation}
\label{m:prime}
m^\prime = \frac{8 \pi r^n}{n} \rho \geq 0 \,.  
\end{equation}
Thus, under the WEC, the generalized Misner-Sharp mass, $m(r)$, is a non-decreasing function of the area radius $r$.

From the regularity at the horizon $r_H$, we can derive a useful constraint on $T^\mu_\nu$. 
We first note that the scalar $T^{\mu}_\nu T_{\mu}^\nu=(T^t_t)^2 + (T^r_r)^2 + T^i_jT^j_i$ is regular on the horizon $r_H$, hence each of the mixed components $T^\mu_\nu$ must also be regular at $r_H$. 
Next, in order to see the relation between each components $T^\mu_\nu$ on the horizon, we introduce the null vector fields defined by 
\begin{equation}
\begin{split}
n^{(\pm)}_\alpha dx^\alpha &= dt \pm \frac{e^\delta}{\mu} dr \,.
\end{split}
\end{equation}
On the horizon $r_H$, the null vector field $n_+^\alpha$ (resp. $n_-^\alpha$) becomes tangent to the past-directed null geodesic generators of the black hole 
horizon at $r_b$ (resp. cosmological horizon at $r_c$). 
Now contracting the EMT with the null vector field $n^\alpha$, we have 
\begin{equation}\label{condi:TttTrr}
{T^{(\rm tot)}}_{\mu \nu} n^\mu n^\nu = \frac{e^{2\delta}}{\mu} \left( {T^{(\rm tot)}}^r_r - {T^{(\rm tot)}}^t_t \right) = \frac{e^{2\delta}}{\mu} \left( {T}^r_r - {T}^t_t \right) \,.
\end{equation}
This quantity should be regular on the horizon $r_H$. Since each component $T^t_t$ and $T^r_r$ must be regular at $r_H$ by itself, as noted above, we find\footnote{
When $k \neq 0$, this can be also shown directly by examining \eqref{conserv_n} as in \cite{Pena:1997cy}.  
One can reexpress \eqref{conserv_n} as 
\begin{equation}
 \dfrac{dE}{dx} = \dfrac{e^{-\delta}}{2r} 
                \left[ 
                      \dfrac{( n - 1 ) k}{ \mu^{1/2}} \left( {T}^t_t - {T}^r_r \right) + \mu^{1/2}\left\{ (3-n) {T}^t_t -(n-1)T^r_r + 2\left( 2n P - {T} \right) \right\} 
                \right] \,. 
\nonumber 
\end{equation}
Here in the left-hand side the derivative is taken with respect to the proper radial distance $x$ defined $dx:= \mu^{-1/2}dr$, which is well-defined on 
the horizon $r_H$, and therefore the right-hand side must also be well-defined on $r_H$. From the first term of the right-hand side, 
one finds \eqref{rho_h_n} when $k\neq 0$. 
}, ${T}^t_t (r_H) = {T}^r_r (r_H)$, namely, 
\begin{equation}
\label{rho_h_n}
\rho(r_H)=-P_r(r_H) \geq 0 \,.  
\end{equation} 

Combining \eqref{divmu_n} and \eqref{divdelta_n} together with the $r$-component of the conservation law $\nabla_\mu {T^{(\rm tot)}}^\mu_{\nu} = 0$, we can derive the following formula:  
\begin{equation}
\label{conserv_n}
\begin{split}
 E' 
&= \frac{e^{-\delta}}{2\mu r} \bigg[ ( n - 1 ) ( k + \mu ) \left( {T}^t_t - {T}^r_r \right) - 2 ( n - 2 )\mu {T}^t_t + 2\mu \left( 2n P - {T} \right) \bigg] \,, 
\end{split}
\end{equation}
where in the right-hand side 
$T$ denotes the trace of ${T}^\mu_\nu$, and in the left-hand side, we have introduced 
\begin{equation}
\label{def:E} 
E \equiv e^{-\delta} {T^{(\rm tot)}}^r_r \,. 
\end{equation}  
This is a higher-dimensional generalization of the formula given by Pena-Sudarsky (see equation (6) in \cite{Pena:1997cy}), and corresponds to the formula for higher-dimensional, spherically symmetric case given by Ghosh-Sk-Sarkar (see equation (2) in~\cite{Ghosh:2023kge}).   
The formula \eqref{conserv_n} can be cast into the form: 
\begin{equation}
\label{divE_n}
E^\prime 
  = \frac{e^{-\delta}}{2\mu r} 
  \left\{ {\cal A} \left( \rho + P \right) - {\cal B} \left( P_r - P \right) \right\} \,,
\end{equation}
where 
\begin{equation}
\label{AB_n}
\mathcal{A} \equiv ( n - 1 ) ( \mu - k ) = ( n - 1 ) \left( - \frac{2 m(r)}{r^{n-1}} - \lambda r^2 \right) ,\quad \mathcal{B} \equiv( n - 1 ) k + ( n + 1 ) \mu \,.
\end{equation}

\section{No-hair properties with or without cosmological constant in $(2+n)$-dimensions}
\label{sec:3}

Let us now turn to the study of no-hair properties of possible static black holes described by the metric \eqref{metric_n} with $n \geq 2$. 
We are concerned with DOC, where $\mu(r)\geq 0$. We analyze the asymptotically flat, de Sitter, and AdS case, separately. 

\subsection{Asymptotically flat case ($\Lambda = 0$)}
\label{zeroCC}

We state our no-hair theorem for static black holes with $\Lambda=0$, which corresponds to a generalization of Pena-Sudarsky's theorem~\cite{Pena:1997cy} to the higher dimensional case.     

\medskip 
\noindent
{\em Theorem~1}. %
{\it Consider a $(2+n)$-dimensional static black hole spacetime $(M, g_{\mu \nu})$ described by the metric \eqref{metric_n} with the metric function~\eqref{lapse_n} and which has a non-degenerate regular black hole horizon at $r_b$ 
and satisfies the Einstein equations with vanishing cosmological constant $\Lambda=0$ and matter fields with $T^{\mu}_{\nu}$. We assume that the generalized Misner-Sharp mass $m(r)$ is positive, \eqref{condi:m}, and asymptotically behaves as \eqref{condi:m:asympt}, and also that the matter fields satisfy the WEC \eqref{wec_n} and asymptotically behave as~\eqref{condi:asympt}. 
If furthermore the following condition between the radial pressure $P_r$ and the tangential pressure $P$:   
\begin{equation}
\label{excon_n}
  P_r \geq P \,, 
\end{equation}
is satisfied, then $T^\mu_\nu$ must vanish identically and $(M,g_{\mu \nu})$ corresponds to the generalized Schwarzschild solution with $({\cal K}_n, k=1)$. 
} 

\medskip 

\noindent
{\it Proof.} For $k=0, -1$ cases, the assumption~\eqref{condi:m} implies $\mu <0$, and therefore the only relevant case is $k=1$. 
In this case, for any $r \geq r_b$, we have ${\cal B}>0$, while ${\cal A} <0$ due to \eqref{condi:m}.  
From WEC and the condition \eqref{excon_n}, it follows that $\rho+P \geq 0$ and $P_r -P \geq 0$. Therefore, \eqref{divE_n} immediately implies $E' \leq 0$; namely that $E$ is a non-increasing function of $r$, thus in particular $E(r_b) \geq E(r=\infty)$. 
By a suitable rescaling of the time coordinate $t$, without loss of generality, one can make $\delta(r_b)=0$. Then, on one hand, we have $E(r_b)= P_r(r_b)=-\rho(r_b) \leq 0$ by the regularity \eqref{rho_h_n} and WEC. On the other hand, by the asymptotic condition~\eqref{condi:asympt}, as well as the regularity $\delta(\infty)<\infty$, the limit $E(\infty) = e^{-\delta(\infty)}P_r(\infty)$ must vanish. 
Therefore $E(r)$ must be identically vanishing $E=0$, thus $P_r=0$, for any $r \geq r_b$. This also implies $E'=0$ identically. 
It then follows from \eqref{divE_n} and \eqref{AB_n} that ${\cal A} \rho + 2n\mu P =0$ for any $r \geq r_b$. This equality holds only when 
$\rho=P=0$ for any $r \geq r_b$, since ${\cal A} <0$ due to \eqref{condi:m}, and the condition~\eqref{excon_n} with $P_r=0$ now implies $P \leq 0$, while $\rho \geq 0$ due to WEC. 
\hfill $\Box$ 

\medskip 

\noindent
{\it Remarks.} 
\begin{itemize}
\item The last part of the statement that the DOC with vanishing $T^\mu_\nu$ must corresponds to the generalized Schwarzschild spacetime in $(2+n)$-dimensions conforms to the uniqueness theorem shown in~\cite{Gibbons:2002av}. 

\item The above condition~\eqref{excon_n} has been employed in \cite{Pena:1997cy}. This is, for example, the case of a static configuration of minimally coupled real scalar field (see also below for time-dependent complex scalar case). Combining~\eqref{excon_n} with the regularity condition~\eqref{rho_h_n} at the horizon, we have, on the horizon, $\rho(r_H)+ P(r_H) \leq 0$. Then, it follows from the WEC that $\rho(r_H)=-P_r(r_H)=-P(r_H)$ and hence
$T^\mu_\nu = - \rho(r_H)  \delta^\mu_\nu$ on the horizon. 
Note that this is not the case for Maxwell fields, in accord with the Reissner-Nordstrom black hole, which is not viewed as a hairy black hole as it has the conserved charge defined at infinity.  

\item As in~\cite{Pena:1997cy}, Theorem~1 can also apply to harmonically time-dependent complex scalar fields~\cite{Lin:2020asf,Matsumoto:2022zvj}, which can be expressed as 
\begin{equation}
 \phi(t,r,z^i)= e^{-i\omega t} \Phi(r) \mathbb{S}(z) \,,   
\end{equation}
where $\Phi(r)$ is a real function and $\mathbb{S}(z)$ denotes the scalar harmonics on ${\cal K}_n$. The EMT for such a complex scalar field is given by 
\begin{equation}
 T_{\mu \nu}(\phi)= \dfrac{1}{2} \left( \nabla_\mu \phi^* \nabla_\nu \phi + \nabla_\mu \phi \nabla_\nu \phi^* \right) 
                       -\dfrac{1}{2} g_{\mu \nu}\left\{ \nabla^\lambda \phi^* \nabla_\lambda \phi + 2V(|\phi|) \right\} \,, 
\end{equation}   
and we can read off  
\begin{eqnarray}
 \rho 
  = \dfrac{\omega^2}{2\mu e^{-2\delta}}\Phi^2 + \dfrac{1}{2} \mu \Phi'{}^2 + V(\Phi) \,, \quad 
  P_r 
   = \rho - 2V(\Phi), \quad 
  P = P_r - \mu \Phi'{}^2 \,. 
\end{eqnarray}  
Then, we can see that $\rho, P_r, P$ in fact satisfy not only WEC but also the requirement~\eqref{excon_n}, and therefore apply Theorem~1. 

\item In Table~\ref{table:1}, we list some known hairy black hole solutions, which all satisfy WEC but violate the condition~\eqref{excon_n} outside the horizon. For comparison, we also include the Reissner-Nordstrom solution, which is not classified as a hairy black hole. See also \cite{Nunez:1996xv} for some other examples of hairy black holes, for which the WEC becomes non-trivial; $\rho + P_r \neq 0$. We also inspect the energy conditions for a one-parameter family of asymptotically flat, static spherically symmetric metrics considered by Johannsen and Psaltis~\cite{Johannsen:2011dh} in Appendix~A.

\medskip 

\begin{table}[h]
\centering
\setlength{\tabcolsep}{0.3pt}
\setlength{\extrarowheight}{8pt} 
\caption{\small Energy conditions and some example of hairy black holes: Reissner-Nordstrom solution (RN), Bardeen solution, Hayward solution~\cite{Hayward:2005gi}, Culetu-Simpson-Visser solution (CSV)~\cite{Culetu:2013fsa,Culetu:2014lca,Simpson:2019mud}, and modified-Culetu-Simpson-Visser solution (mCSV)~\cite{Xiang:2013sza}, Dymnikova solution~\cite{Dymnikova:1992ux} and Kiselev solution~\cite{Kiselev:2002dx}. In the table, $M$ and $Q$ correspond to the asymptotic (global) charges (ADM mass and electric charge, respectively), and $L, a, r_0, w_n, r_n$ to additional parameters, characterizing the corresponding hairy black hole solutions. Note that for Kiselev solution, the state parameter $w_n$ of quintessence is assumed to be in the range $-1<w_n<0$. They all satisfy WEC but violate the condition~\eqref{excon_n} outside the horizon.  
}

{\small 
\label{table:1}
\begin{tabular}{l|c|c|c|c|c}\hline
  &  & \multicolumn{3}{c|}{{WEC}}  & {Condition~\eqref{excon_n}} \\[8pt]\hline
  
  {Solutions} & & $ 8\pi \rho $ & ${ }\, 8\pi (\rho + P_r) \, $ & $ 8\pi (\rho + P) $ & $ 8\pi (P_r- P) $ \\[8pt] \hline \hline
  
  RN &  & $ \dfrac{Q^2}{r^4} $ & $ 0 $ & $ \dfrac{2Q^2}{r^4} $ & $ -\dfrac{2Q^2}{r^4} $ \\[9pt]\hline
  
  
  Bardeen & & $ \dfrac{6MQ^2}{r^5} $ & $ 0 $ & $ \dfrac{15MQ^2}{r^5} $ & $ -\dfrac{15MQ^2}{r^5} $ \\[9pt]\hline
  
  
  Hayward && $ \dfrac{6ML^3}{(r^3+L^3)^2} $ & $ 0 $ & $ \dfrac{18ML^3r^3}{(r^3+L^3)^3} $ &  $ -\dfrac{18ML^3r^3}{(r^3+L^3)^3} $ \\[9pt]\hline
  
  CSV & & $ \dfrac{2Ma}{r^4}e^{-{a}/{r}} $ & $ 0 $ & $ \dfrac{Ma(4r-a)}{r^5}e^{-{a}/{r}} $ & $ -\dfrac{Ma(4r-a)}{r^5}e^{-{a}/{r}} $
  \\[9pt]\hline
  
  mCSV & & $ \dfrac{4Ma^2}{r^5}e^{-{a^2}/{r^2}} $ & $ 0 $ & $ \dfrac{2Ma^2(5r^2-2a^2)}{r^7}e^{- {a^2}/{r^2}} $ & $ -\dfrac{2Ma^2(5r^2-2a^2)}{r^7}e^{- {a^2}/{r^2}} $
  \\[9pt]\hline

  Dymnikova & &
  $ \dfrac{3}{r_0^2}e^{-\frac{r^3}{2Mr_0^2}} $
  & $ 0 $ 
  & $ \dfrac{9r^3}{4Mr_0^4}e^{-\frac{r^3}{2Mr_0^2}} $ 
  & $ -\dfrac{9r^3}{4Mr_0^4}e^{-\frac{r^3}{2Mr_0^2}} $  
  \\[9pt]\hline

   Kiselev & &
   $ -\dfrac{3w_n}{r^2}\left(\dfrac{r_n}{r}\right)^{3w_n+1}  $
  & $ 0 $ 
  & $ -\dfrac{9w_n(w_n+1)}{2r^2}\left(\dfrac{r_n}{r}\right)^{3w_n+1}  $ 
  & $ \dfrac{9w_n(w_n+1)}{2r^2}\left(\dfrac{r_n}{r}\right)^{3w_n+1} $ 
  \\[9pt]\hline

\end{tabular}
} 
\end{table}

\end{itemize}

\bigskip 

We also have the following theorem concerning ``hair length" as a generalization of Nunez-Quevedo-Sudarsky's no-short-hair theorem~\cite{Nunez:1996xv} to higher dimensions.   

\bigskip 

\noindent
{\em Theorem~2}.  
{\it Consider the class of static black hole spacetimes considered in Theorem~1. We assume that the generalized Misner-Sharp mass $m(r)$ is positive, \eqref{condi:m}, and asymptotically behaves as \eqref{condi:m:asympt} and also that the matter fields satisfy WEC, \eqref{wec_n}, as in Theorem~1. We further require that $r^{n+2}E$ vanishes at infinity. If furthermore the trace of the EMT is negative, then $r^{n+2}E$ is negative semidefinite at the horizon $r_b$ and is decreasing between $r_b$ and $r_0$, where $r_0 > r_\star :=[(n+1)/2]^{1/(n-1)}r_b$ and for some $r> r_0$, the function $r^{n+2}E$ begins to increase toward its asymptotic value, 0.  
}

\medskip 

\noindent
{\it Proof.} 
As shown in the proof of Theorem~1, only for the case $k=1$, the metric \eqref{metric_n} with~\eqref{lapse_n} admits a static black hole.  
From \eqref{divE_n} we can obtain the following formula
\begin{equation}
\label{div:rE_n}
(r^{n+2}E)^\prime = \frac{r^{n+1}e^{-\delta}}{2\mu } 
  \left\{ {\cal C} \left( \rho + P_r \right) + 2 \mu T \right\} \,,
\end{equation}
where ${\cal C}:= (n+1)\mu - (n-1)$ and $T$ denotes the trace of EMT. 
As has already been shown in the proof of Theorem~1, $E(r_b) \leq 0$, hence $r_b^{n+2}E(r_b) \leq 0$.  
By WEC $\rho+P_r \geq 0$ and by assumption $T < 0$, we find that $(r^{n+2}E)^\prime < 0 $, and hence $r^{n+2}E$ is a decreasing function, unless ${\cal C}>0$. 
Since at the horizon $r_b$, the regularity implies that $(\rho +P_r)/\mu $ is finite, and ${\cal C}(r_b)=-(n-1)<0$, hence $(r^{n+2}E)^\prime \leq 0$ at $r_b$. Thus, $r^{n+2}E$ is a negative, decreasing function between $r_b$ and some $r_0>r_1$, where $r_1$ is given by ${\cal C}(r_1)=0$. 
By straightforward calculation, one finds $r_1^{n-1} = (n+1)m(r_1)$. Since the generalized Misner-Sharp mass is non-decreasing 
as shown by \eqref{m:prime}, it follows that $r_0^{n-1} >r_1^{n-1}=(n+1)m(r_1) \geq (n+1)m(r_b) =(n+1)r_b^{n-1}/2$.   
\hfill $\Box$ 

\bigskip 

\noindent
{\it Remarks.} 
\begin{itemize}
\item The requirement that $r^{n+2}E(=r^{n+2}e^{-\delta}P_r)$ goes to zero at infinity $r \rightarrow \infty$ may be somewhat more stringent condition than our asymptotic condition~\eqref{condi:asympt}, in particular, than the standard definition of asymptotic flatness in $4$-dimensions. For example, $r^4e^{-\delta}P_r$ for Maxwell fields in $4$-dimensions remains to be non-vanishing at infinity, as pointed out in~\cite{Nunez:1996xv}. Note however that in higher dimensions, Maxwell fields satisfy this requirement.  

\item 
From Theorem~2, we have $ r_\star^{n+2} E(r_\star) <r_b^{n+2}E(r_b) \leq 0$, which is rewritten in terms of the radial pressure $P_r<0$ as  
\begin{equation}
\label{ineq:Pr}  
|P_r(r_\star)| > \left(\dfrac{r_b}{r_\star}\right)^{n+2}|P_r(r_b)| e^{\delta(r_\star)- \delta(r_b)} \,. 
\end{equation} 
By integrating \eqref{divdelta_n}, we have
\begin{equation}
 \delta(r_\star) - \delta (r_b) = - \dfrac{8\pi}{n} \int_{r_b}^{r_\star} dr \dfrac{r(\rho+P_r)}{\mu} \,.
\end{equation} 
This provides a bound on $\delta(r_\star)-\delta(r_b)$ in terms of the maximal value of the matter fields in the interval $[r_b, r_\star]$. 
Substituting this into the right-hand side of \eqref{ineq:Pr}, we can, in principle, obtain a bound on the radial pressure $P_r$ at the point $r_\star$. Following a similar argument of~\cite{Nunez:1996xv}, we can claim that a black hole hair, if exists, must extend beyond the point $r_\star$. When $n=2$, the lower bound $r_\star :=[(n+1)/2]^{1/(n-1)}r_b$ reduces to the universal lower bound $(3/2)r_b$ introduced in the original no-short hair conjecture~\cite{Nunez:1996xv} for $4$-dimensional black holes. 

\item We also note that the lower bound $r_\star :=[(n+1)/2]^{1/(n-1)}r_b$ coincides precisely with the radius of the photon-sphere (i.e., the radius of circular null geodesics) for the $(2+n)$-dimensional Schwarzschild black hole. In this respect, it is intriguing to note that as discussed by Hod~\cite{Hod:2011aa} and also by Ghosh-Sk-Sarkar~\cite{Ghosh:2023kge}, the hair length may be bounded by the photon-sphere radius $r_\gamma$: One can claim that non-trivial hair, if exists, must extend the photon-sphere. We discuss this point in more detail in Appendix~B. 

\end{itemize}

\subsection{Asymptotically de Sitter case ($\Lambda >0$)}
\label{posCC}

For $k=0,-1$, we have $\mu \leq 0$ due to \eqref{condi:m} and the positivity $\lambda >0$.  
Hence the only relevant case is $k=1$, for which it immediately follows from \eqref{AB_n} that ${\cal A}<0$ while ${\cal B} > 0$ in DOC. Then, with WEC and the condition \eqref{excon_n}, we find that $E' \leq 0$, namely that $E$ is a non-increasing function of $r$ in DOC, thus in particular $E(r_b) \geq E(r_c)$. 
Note that as in the asymptotically flat case, one can rescale the time coordinate $t$ so that $\delta (r_b) = 0$. It follows from \eqref{divdelta_n} and WEC that $\delta(r)$ becomes a non-increasing function in DOC, and hence, $e^{-\delta(r_c)} \geq e^{-\delta(r_b)}=1$. 
Now let us introduce the non-negative constant $\epsilon := [{n(n+1)}/{16\pi}] \lambda \left( e^{-\delta (r_c)} - 1 \right) \geq 0$. Then, we obtain the following theorem for static black holes with positive cosmological constant:  

\medskip 
\noindent
{\em Theorem~3}.  
{\it Consider a $(2+n)$-dimensional static black hole spacetime $(M, g_{\mu \nu})$ described by the metric \eqref{metric_n} which satisfies the Einstein equations with a positive cosmological constant $\Lambda>0$ and which is bounded by two non-degenerate regular Killing horizons, i.e., a black hole horizon at $r_b$ and a cosmological horizon at $r_c> r_b$. 
If furthermore, the positivity~\eqref{condi:m} of the generalized Misner-Sharp mass $m(r)$ holds and the matter fields satisfy the WEC \eqref{wec_n} and the conditions \eqref{excon_n}, then the matter fields must satisfy the following condition:  
\begin{equation}
\begin{split}
\label{condi:hair}
\rho(r_b) &\leq \rho(r_c) e^{-\delta (r_c)} + \epsilon \,. 
\end{split}
\end{equation}
} 

\bigskip 

\noindent
{\it Proof.} By using the definition \eqref{def:E} of $E$ as well as the regularity condition \eqref{rho_h_n}, the condition $E(r_b) \geq E(r_c)$ can be expressed in terms of the energy density $\rho$ of the matter field at the horizons, resulting in~\eqref{condi:hair}. 
\hfill $\Box$

\bigskip 

\noindent
{\it Remarks.} 
\begin{itemize}

\item Although the inequality~\eqref{condi:hair} is a non-trivial condition due to the factor $\delta(r_c)$ and $\epsilon$, it can roughly be interpreted that in order for a static black hole to have a non-trivial hair stemming from $T^\mu_\nu$, the matter field should not decay so quickly toward the cosmological horizon 
to meet \eqref{condi:hair}. In other words, the hair of a static black hole, if exists under WEC and \eqref{excon_n}, must be long enough so that it reaches the cosmological horizon. In this sense, Theorem~3 above may be viewed as a generalization of the no-short hair theorem of \cite{Nunez:1996xv} to include a positive cosmological constant. In this respect, note that a no-short-hair theorem in $4$-dimensional asymptotically (anti-)de Sitter spacetimes has also been discussed in~\cite{Cai:1997ij}. Note also that some known example, e.g., MTZ solution~\cite{Martinez:2002ru} and BdS solution~\cite{Fernando:2016ksb} violate the condition~\eqref{condi:hair} but have hair since they violate the condition~\eqref{excon_n}. 

\item When $\Lambda$ vanishes, $r_c \rightarrow \infty$, $\rho(r_c) \rightarrow 0$ for the asymptotic condition, the condition~\eqref{condi:hair} indicates a possible violation of WEC. 

\item Theorem~3 indicates that if the condition~\eqref{condi:hair} is not satisfied, then $T^\mu_\nu$ would vanish identically in DOC and such a black hole $(M,g_{\mu \nu})$ should correspond to the static region of a generalized Schwarzschild-de Sitter spacetime with $({\cal K}_n, k=1)$. In this respect, we note that a uniqueness theorem for Schwarzschild-de Sitter spacetime in $4$-dimensions ($n=2$) has recently been shown \cite{MaAYu15,Borghini:2019msu}. 

\end{itemize}

\subsection{Asymptotically AdS case ($\Lambda <0$)}
\label{negCC}

In the asymptotically AdS case, we have the following no-hair theorem, which generalizes Sudarsky-Gonzalez's theorem~\cite{Sudarsky:2002mk} to higher-dimensional case. 

\medskip 
\noindent
{\em Theorem~4}.  
{\it Consider a $(2+n)$-dimensional static black hole spacetime $(M, g_{\mu \nu})$ described by the metric \eqref{metric_n} and which has a non-degenerate regular black hole horizon at $r_b$ and satisfies the Einstein equations with a negative cosmological constant $\Lambda<0$ and matter fields with $T^{\mu}_{\nu}$. We assume that the generalized Misner-Sharp mass $m(r)$ is positive, \eqref{condi:m}, and asymptotically behaves as \eqref{condi:m:asympt}, and also that the matter fields satisfy the WEC \eqref{wec_n} and asymptotically behave as~\eqref{condi:asympt}. 
If further the matter fields satisfy the following conditions, 
\begin{eqnarray}
\label{condi:nohair:ads}
 \rho + P_r \leq \dfrac{2n}{(n-1)}(P_r-P) \,, 
\end{eqnarray}
then there are no nontrivial static black hole solutions with $k=0,1$ in the asymptotically AdS spacetime whose asymptotic behavior corresponds to the AdS spacetime with the true cosmological constant $\lambda=2\Lambda/n(n+1)<0$. 
} 

\bigskip 

\noindent
{\it Proof.} The formula~\eqref{conserv_n} can be cast into the following form, 
\begin{eqnarray}
\label{div:Pr}
(e^{-\delta}P_r)^\prime 
  &=& \dfrac{ e^{-\delta} }{2\mu r} 
  \left[ 
         -\{(n-1)k+ (n+1)(-\lambda) r^2 \} \left( \rho + P_r \right) 
    \right.
       \nonumber \\
  &{}& \qquad 
   \left. 
         + \mu \left\{ (n-1)(\rho+P_r) -2n(P_r- P) \right\} 
  \right] \,. 
\end{eqnarray}
It is straightforward to see that when $k=0, 1$, the first term of the right-hand side of \eqref{div:Pr} is always non-positive as $\lambda<0$ and WEC. The condition~\eqref{condi:nohair:ads} ensures the non-positivity of the the second term of the right-hand side of \eqref{div:Pr}. Then, $e^{-\delta}P_r$ is a non-increasing function in $r \geq r_b$, and we can derive a contradiction by similar argument in the proof of Theorem~1. \hfill $\Box$

\bigskip

\noindent
{\it Remarks.} 
\begin{itemize}
\item When $\rho+P_r = P_r-P$, the condition~\eqref{condi:nohair:ads} holds in any dimensions $n \geq 2$.  
This is, for example, the case of a minimally coupled scalar field, 
as studied in $4$-dimensional spherical black hole case~\cite{Sudarsky:2002mk}. 

\item 
If, for example, a scalar field admits non-vanishing potential $V(\phi)$ at infinity, 
namely, if it does not satisfy our asymptotic conditions, then a non-trivial hairy static black hole is allowed, in which the effective cosmological constant $\Lambda_{\rm eff}:= \Lambda + 8\pi V(\phi_{\rm \infty})$ determines its asymptotic structure, as explained in~\cite{Sudarsky:2002mk}. 

\item For static, vacuum spacetimes with a negative cosmological constant, uniqueness theorems have been proposed with some further assumptions (see, e.g., \cite{Boucher:1983cv, Anderson:2002xb}). 
\end{itemize}

\section{Summary}
\label{sec:4} 
We have shown the four theorems concerning no-hair and no-short hair properties of static black holes in four and higher dimensional asymptotically locally flat, de Sitter, and AdS spacetimes. 
Theorems~1 and~4 correspond, respectively, to higher-dimensional generalizations of the no-hair theorem by Pena-Sudarsky~\cite{Pena:1997cy} 
and that by Sudarsky-Gonzalez~\cite{Sudarsky:2002mk} with a negative cosmological constant. Theorems~2 and~3 correspond, respectively, to higher-dimensional generalizations of the no-short hair theorem by 
Nunez-Quevedo-Sudarsky~\cite{Nunez:1996xv} for asymptotically flat case and that by Cai-Ji~\cite{Cai:1997ij} with a positive cosmological constant. 
All our theorems assume that the static metric takes the warped product form of \eqref{metric_n}, admits regular non-degenerate Killing horizons as a black hole event horizon and also a cosmological horizon for $\Lambda>0$ case, the Misner-Sharp local mass function is positive definite, and matter fields satisfy WEC. Then, each theorem further requires its own additional assumption, such as \eqref{excon_n} in Theorem~1 and the negative trace of energy-momentum tensor in Theorem~2. As shown in Table~\ref{table:1}, some known hairy black hole solutions satisfy WEC but violate our additional condition~\eqref{excon_n}.

Our metric ansatz of the warped product type~(\ref{metric_n}) is, on one hand, general enough to encompass a fairly large class of static black holes as explained below eq.~(\ref{metric_n}). On the other hand, it is restrictive enough so as to make possible to reduce our analyses to essentially an analysis in the $2$-dimensional spacetime spanned by $(t, r)$, just like the case of the $4$-dimensional theorems. As a result, the parameter $n$ that specifies the spacetime dimension comes in our proofs in rather a straightforward manner, except the requirement of the asymptotic behavior of matter fields, eq.~(\ref{condi:asympt}). In fact, the formula~(\ref{condi:TttTrr})---which relates $T^t_t$ and $T^r_r$ and derives the regularity at the horizon---is independent of $n$. 
Also thanks to the ansatz~(\ref{metric_n}), the possible geometry of the $n$-dimensional space ${\cal K}_n$ can be fixed throughout the domain of outer communications. This is, in general, no longer the case in higher dimensions once we consider a more general, non-factorizable form of the metric as recently numerically demonstrated~\cite{Horowitz:2022hlz}.  

The no-short hair theorems~2 and 3 above imply that if a black hole has a hair, one can, in principle, measure the non-trivial structure of matter fields producing the hair at, at least, the lower bound radius $r_\star$ of the hairy region (or the photon-sphere radius $r_\gamma$ as shown in Appendix). In other words, it may be possible to define a set of quasi-local quantities at $r_\star$ in such a way that they can contain essential information about the nature of the black hole hair. 
Consider, for example, the hairy black holes listed in Table~\ref{table:1} (apart from the Reissner-Nordstrom solution and Kiselev solution), which possess the two parameters: $M$ and one of $Q$, $L$, $a$, or $r_0$. The asymptotic charge (the ADM mass) specifies only $M$ but the Misner-Sharp mass $m(r_\star)$ quasi-locally defined at $r_\star$ can be used to specify the other parameter (either $Q$, $L$, $a$, or $r_0$). In more general cases (e.g., Kiselev solution and some examples listed in~\cite{Nunez:1996xv}), one may need more, different types of quasi-local quantities, other than the Misner-Sharp mass, in order to completely characterize hairy black holes. It would be interesting to clarify whether or not (if possible, under what circumstances) one can uniquely specify the character of hairy black holes in terms of a set of quasi-local charges at $r_\star$ (or $r_\gamma$), besides the asymptotic conserved charges.

In the present paper, for simplicity we have assumed that the Killing horizons are non-degenerate ones. There seems to be no serious obstruction to generalize the present four theorems to involve degenerate horizons. It is worth studying whether we can further relax our conditions to obtain similar theorems for extremal black holes.  

In the AdS/CFT correspondence, hairy black holes have played an important role. It has been revealed that some unstable charged hairy black holes in AdS spacetimes are dual to the boundary superconductor~\cite{Hartnoll:2008vx}. Instabilities of hairy black holes in the bulk give rise to phase transitions in the AdS conformal boundary. In this context, our no-hair theorem in the asymptotically AdS case may be used to place some non-trivial restrictions on possible models of the holographic superconductor.

\bigskip
\goodbreak
\centerline{\bf Acknowledgments}

\medskip 
\noindent
This work was supported in part by JSPS KAKENHI Grant No. 20K03938 and also supported by MEXT KAKENHI Grant-in-Aid for Transformative Research Areas A Extreme Universe No. 21H05182, 21H05186. We are grateful to the long term workshop YITP-T-23-01 held at YITP, Kyoto University, where a part of this work was done.

\section*{Appendix A: WEC and the condition $P_r \geq P$ for Johannsen-Psaltis parameterized metrics}\label{app:A}

We examine WEC and the condition~\eqref{excon_n} (i.e., $P_r \geq P$) for a one-parameter family of the asymptotically flat, static, spherically symmetric metrics, obtained by taking the non-rotating limit of the Johannsen and Psaltis~\cite{Johannsen:2011dh}: 
\begin{equation}
\label{metric_jp}
ds^2 = - \left( 1 - \frac{2M}{r} \right) \left( 1 + \frac{\epsilon_3 M^3}{r^3} \right) dt^2 + \left( 1 - \frac{2M}{r} \right)^{-1} \left( 1 + \frac{\epsilon_3 M^3}{r^3} \right) dr^2 + r^2 d\Omega^2 ,
\end{equation}
with mass $M$ and dimensionless parameter $\epsilon_3$. When $\epsilon_3 =0$ this metric reduces to the Schwarzschild metric. 
To investigate the WEC and the condition~\eqref{excon_n} in this spacetime, we calculate the Einstein tensor and its combinations: 
\begin{equation}
\label{wec_jp}
\begin{split}
-G^t_t &= - \frac{\epsilon_3 M^3 \left( 2r^3 - 6Mr^2 - \epsilon_3 M^3 \right)}{r^2 \left( r^3 + \epsilon_3 M^3 \right)^2} \,, 
\\
G^r_r - G^t_t &= - \frac{6 \epsilon_3 M^3 \left( r - 2M \right)}{\left( r^3 + \epsilon_3 M^3 \right)^2}\,, 
\\
G^\theta_\theta - G^t_t &= \frac{\epsilon_3 M^3 \left( 8r^6 - 18Mr^5 + \epsilon_3 M^3 r^3 + 2\epsilon_3^2 M^6 \right)}{2r^2 \left( r^3 + \epsilon_3 M^3 \right)^3} \,,  
\end{split}
\end{equation}
and
\begin{equation}
\label{cond_jp}
G^r_r - G^\theta_\theta = - \frac{\epsilon_3 M^3 \left( 20r^6 - 42Mr^5 + 13\epsilon_3 M^3 r^3 - 24\epsilon_3 M^4 r^2 + 2\epsilon_3^2 M^6 \right)}{2r^2 \left( r^3 + \epsilon_3 M^3 \right)^3} \,. 
\end{equation}
On the assumption of $M>0$, this spacetime admits a regular horizon at $r_s = 2M$ when $\epsilon_3 > -8$. 
One can also find that when $\epsilon_3 <0$, the metric admits a curvature singularity at the finite radius $r_\epsilon := |\epsilon_3|^{1/3} M$ 
and in particular that when $\epsilon_3 \leq -8$, the curvature singularity becomes naked in the sense of $r_\epsilon \geq r_s$. Therefore in what follows we focus on the case $\epsilon_3 >-8$ and $\epsilon_3 \neq 0$. 

First, we examine the conditions for $\epsilon_3 > 0$. Although $-G^t_t$ is positive at the horizon $r_s$: 
\begin{equation}
\label{wec1_jp}
- G^t_t (r_s) = \frac{\epsilon_3}{4M^2 ( \epsilon_3 + 8 )} > 0 \,, 
\end{equation} 
it becomes negative at asymptotic region $r \rightarrow \infty$, implying the violation of the WEC, $\rho \geq 0$, at large distances. As we can see from \eqref{wec_jp}, $G^r_r - G^t_t$ vanishes at $r = r_s$ and becomes negative outside the horizon $r_s$ where the condition $\rho + P_r \geq 0$ is always violated. $G^\theta_\theta - G^t_t$ on the horizon is expressed as 
\begin{equation}
\label{wec3_jp}
G^\theta_\theta (r_s) - G^t_t (r_s) = \frac{\epsilon_3 ( \epsilon_3 - 4 )}{4M^2 ( \epsilon_3 + 8 )^2} \,. 
\end{equation}
Thus, \eqref{wec3_jp} becomes positive semidefinite for $\epsilon_3 \geq 4$, while negative for $0 < \epsilon_3 < 4$. In the asymptotic region $r \rightarrow \infty$, we find that $G^\theta_\theta - G^t_t$ approaches 0 from above. Therefore, the condition $\rho + P \geq 0$ is violated near the horizon for $0 < \epsilon_3 < 4$, while for $\epsilon_3 \geq 4$, the condition $\rho + P \geq 0$ is satisfied if it has no zero outside the horizon. 
$G^r_r - G^\theta_\theta$ is written as
\begin{equation}
G^r_r (r_s) - G^\theta_\theta (r_s) = - \frac{\epsilon_3 ( \epsilon_3 - 4 )}{4M^2 ( \epsilon_3 + 8 )^2} \,,
\end{equation}
on the horizon. This quantity corresponds to \eqref{wec3_jp} with reversed sign, and hence becomes negative semidefinite for $\epsilon_3 \geq 4$, while positive for $0 < \epsilon_3 < 4$. In the asymptotic region $r \rightarrow \infty$, $G^r_r - G^\theta_\theta$ approaches 0 from below. Hence, for $0 < \epsilon_3 < 4$, the condition $P_r \geq P$ is violated at large distances from the horizon, and it may always be violated for $\epsilon_3 \geq 4$.   

Next, we consider the conditions for $-8< \epsilon_3 < 0$. 
It is found that $-G^t_t$ is negative on the horizon $r_s$ and approaches zero from above in the asymptotic region. Thus $\rho \geq 0$ is violated near the horizon. $G^r_r - G^t_t$ has a zero on the horizon and positive outside of it, so $\rho + P_r \geq 0$ is satisfied. $G^\theta_\theta - G^t_t$ becomes positive on the horizon and asymptotically approaches zero from below. Therefore, $\rho + P \geq 0$ is violated at large distances from the horizon. $G^r_r - G^\theta_\theta$ becomes negative on the horizon and asymptotically approaches zero from above. Thus the condition $P_r \geq P$ is violated near the horizon. We summarize these results in Table~\ref{JPsolution}. 

\medskip 

\begin{table}[h]
\caption{Horizon radii and the range of $r$ where WEC and the condition~\eqref{excon_n} are satisfied for the Johannsen-Psaltis metics with various values of $\epsilon_3$.}
\label{JPsolution}

\medskip 
\begin{tabular}{c|c|c|c|c|c} \hline 
values of $\epsilon_3$	& horizon radius					& $\rho \geq 0$							& $\rho + P_r \geq 0$				& $\rho + P \geq 0$	& $P_r \geq P$	\\
\hline \hline
$4 \leq \epsilon_3$				& \multirow{2}{*}{$r_s$}		& \multirow{2}{*}{$r \sim r_s$}	& \multirow{2}{*}{not satisfied}	& $r \geq r_s$				& not satisfied	\\
$0 < \epsilon_3 < 4$		& 											& 													& 													& $r \gg r_s$				& $r \sim r_s$		\\
\hline
$-8 < \epsilon_3 < 0$		& $r_s , r_\epsilon < r_s$	& $r \gg r_s$								& $r \geq r_s$								& $r \sim r_s$				& $r \gg r_s$		\\
\hline 
\end{tabular}
\end{table}

\section*{Appendix B: No-short hair and null circular orbits } \label{app:B}
In this appendix we consider some close relationship between the limit of short hair and null circular geodesics, following the arguments given by~\cite{Hod:2011aa,Ghosh:2023kge}. 
For simplicity, we assume that our $n$-dimensional manifold $({\cal K}_n,\gamma_{ij})$ admits at least a Killing vector field $\varphi^\mu:= (\partial/\partial \varphi)^\mu, \: \varphi \in z^i$ whose orbits are closed in $({\cal K}_n,\gamma_{ij})$. (Note that this is not always the case in higher dimensions, $n \geq 3$, even if ${\cal K}_n$ is compact.) We then focus on the closed orbits of $\varphi^\mu$ along which apart from the Killing parameter $\varphi$, no other angle coordinates $z^i$ change, so that the lagrangian for causal geodesics reduces to  
\begin{equation}
\label{geodesic:lagrangian}
{L} =\dfrac{1}{2}\left( - \mu e^{-2\delta} \dot{t}^2 + \mu^{-1} {\dot r}^2 + r^2 {\dot \varphi}^2  \right) = -\frac{\epsilon}{2} \,, 
\end{equation}
where the {\em dot} denotes the derivative by an affine parameter and $\epsilon =1$ for a timelike geodesic and $\epsilon =0$ for null geodesic. With respect to the stationary Killing field $\partial /\partial t$ and the circular Killing field $\partial/\partial \varphi$, we find two constants of motion, ${\cal E}:= \mu e^{-2\delta } {\dot t}$ and ${\cal L} := r^2 {\dot \varphi}$. Then, we have
\begin{equation}
 {\dot r}^2 = \mu\left( \dfrac{{\cal E}^2}{\mu e^{-2\delta}}-\dfrac{{\cal L}^2}{r^2}-\epsilon \right) \,.
\end{equation}
The circular orbits are given under the condition 
\begin{equation}
\label{condi:circle}
{\dot r}^2=0, \quad ({\dot r}^2)'=0 \,. 
\end{equation}
Then, for timelike geodesics $\epsilon =1$, we obtain
\begin{equation}
{\cal E}^2= \dfrac{2(\mu e^{-2\delta})^2}{ 2\mu e^{-2\delta} - r(\mu e^{-2\delta})'} , \quad {\cal L}^2 = \dfrac{r^3(\mu e^{-2\delta})'}{ 2\mu e^{-2\delta} - r(\mu e^{-2\delta})'} \,.
\end{equation}
Substituting \eqref{divmu_n} and \eqref{divdelta_n} into the above expressions, we find that the requirement that the test particle's energy ${\cal E}$ is real implies 
\begin{equation}
\label{def:C}
 {\cal C} \geq -(n+1) \lambda r^2 + \dfrac{16\pi}{n}r^2 P_r \,, 
\end{equation}
where ${\cal C}:= (n+1)\mu -(n-1)k$ as considered in Theorem~2. 

As for a circular null geodesics, from the condition~\eqref{condi:circle} with $\epsilon=0$, we find the radius $r_\gamma$ of null circular orbits or the photon-sphere to be determined via  
\begin{equation}
\label{def:rgmm}
 r_\gamma = \dfrac{2 \mu e^{-2\delta}}{(\mu e^{-2\delta})' } \,. 
\end{equation}
Note that for the $(2+n)$-dimensional Schwarzschild black hole case (i.e., $k=1$, $\delta = 0$, and $\mu = 1-(r_b/r)^{n-1}$), 
we can immediately find from (\ref{def:rgmm}) that 
\begin{equation} 
r_\gamma = \left[ \dfrac{n+1}{2}\right]^{1/(n-1) }r_b \,, 
\end{equation}
which precisely coincides with the lower bound $r_\star$ introduced in Theorem~2. 

Now let us define $Q:= e^\delta r^{n+2}E$. Then, combining \eqref{div:rE_n}, \eqref{def:C} and \eqref{def:rgmm}, we find 
\begin{equation}
Q'(r_\gamma) = r_\gamma^{n+1}T^{({\rm tot})} \leq 0 . 
\end{equation}
From \eqref{def:C}, circular geodesics are excluded for ${\cal C}<0$, implying 
\begin{equation}
 Q' (r<r_\gamma) \leq 0 \,. 
\end{equation}
Since on the horizon, $E(r_E)<0$, hence $Q$ is also negative at $r_H$. Therefore, as in the case of $E$, $Q$ is also a negative, non-increasing function between $r_H<r<r_\gamma$. Following the argument of~\cite{Hod:2011aa}, we define $r_{\rm hair}$ to be the point where $|Q(r)|$ attains its local maximum, and then we find the lower bound, 
\begin{equation}
 r_{\rm hair} \geq r_\gamma \,.  
\end{equation} 
Thus, one can claim that the non-trivial hair, if exists, must extend beyond the photon-sphere at $r_\gamma$.


\begin{thebibliography}{99}

\bibitem{Israel:1967wq}
W.~Israel,
Phys. Rev. \textbf{164}, 1776-1779 (1967)
doi:10.1103/PhysRev.164.1776

\bibitem{Israel:1967za}
W.~Israel,
Commun. Math. Phys. \textbf{8}, 245-260 (1968)
doi:10.1007/BF01645859

\bibitem{Carter:1971zc}
B.~Carter,
Phys. Rev. Lett. \textbf{26}, 331-333 (1971)
doi:10.1103/PhysRevLett.26.331

\bibitem{Robinson:1974nf}
D.~C.~Robinson,
Phys. Rev. D \textbf{10}, 458-460 (1974)
doi:10.1103/PhysRevD.10.458

\bibitem{Robinson:1975bv}
D.~C.~Robinson,
Phys. Rev. Lett. \textbf{34}, 905-906 (1975)
doi:10.1103/PhysRevLett.34.905

\bibitem{Mazur:1982db}
P.~O.~Mazur,
J. Phys. A \textbf{15}, 3173-3180 (1982)
doi:10.1088/0305-4470/15/10/021

\bibitem{Chrusciel:2012jk}
P.~T.~Chrusciel, J.~Lopes Costa and M.~Heusler,
Living Rev. Rel. \textbf{15}, 7 (2012)
doi:10.12942/lrr-2012-7
[arXiv:1205.6112 [gr-qc]].

\bibitem{Bekenstein:1971hc}
J.~D.~Bekenstein,
Phys. Rev. D \textbf{5}, 1239-1246 (1972)
doi:10.1103/PhysRevD.5.1239

\bibitem{Pena:1997cy}
I.~Pena and D.~Sudarsky,
Class. Quant. Grav. \textbf{14}, 3131-3134 (1997)
doi:10.1088/0264-9381/14/11/013

\bibitem{Nunez:1996xv}
D.~Nunez, H.~Quevedo and D.~Sudarsky,
Phys. Rev. Lett. \textbf{76}, 571-574 (1996)
doi:10.1103/PhysRevLett.76.571
[arXiv:gr-qc/9601020 [gr-qc]].

\bibitem{Lin:2020asf}
K.~Lin, S.~Zhang, C.~Zhang, X.~Zhao, B.~Wang and A.~Wang,
Phys. Rev. D \textbf{102}, no.2, 024034 (2020)
doi:10.1103/PhysRevD.102.024034
[arXiv:2004.04773 [gr-qc]].

\bibitem{Matsumoto:2022zvj}
S.~Matsumoto,
Class. Quant. Grav. \textbf{40}, no.17, 175011 (2023)
doi:10.1088/1361-6382/ace94e
[arXiv:2210.03966 [gr-qc]].

\bibitem{Bizon:1990sr}
P.~Bizon,
Phys. Rev. Lett. \textbf{64}, 2844-2847 (1990)
doi:10.1103/PhysRevLett.64.2844

\bibitem{Garfinkle:1990qj}
D.~Garfinkle, G.~T.~Horowitz and A.~Strominger,
Phys. Rev. D \textbf{43}, 3140 (1991)
[erratum: Phys. Rev. D \textbf{45}, 3888 (1992)]
doi:10.1103/PhysRevD.43.3140

\bibitem{Bizon:1992gb}
P.~Bizon and T.~Chmaj,
Phys. Lett. B \textbf{297}, 55-62 (1992)
doi:10.1016/0370-2693(92)91069-L

\bibitem{Lavrelashvili:1992ia}
G.~V.~Lavrelashvili and D.~Maison,
Nucl. Phys. B \textbf{410}, 407-422 (1993)
doi:10.1016/0550-3213(93)90441-Q

\bibitem{Greene:1992fw}
B.~R.~Greene, S.~D.~Mathur and C.~M.~O'Neill,
Phys. Rev. D \textbf{47}, 2242-2259 (1993)
doi:10.1103/PhysRevD.47.2242
[arXiv:hep-th/9211007 [hep-th]].

\bibitem{Nucamendi:1995ex}
U.~Nucamendi and M.~Salgado,
Phys. Rev. D \textbf{68}, 044026 (2003)
doi:10.1103/PhysRevD.68.044026
[arXiv:gr-qc/0301062 [gr-qc]].

\bibitem{Herdeiro:2018wub}
C.~A.~R.~Herdeiro, E.~Radu, N.~Sanchis-Gual and J.~A.~Font,
Phys. Rev. Lett. \textbf{121}, no.10, 101102 (2018)
doi:10.1103/PhysRevLett.121.101102
[arXiv:1806.05190 [gr-qc]].

\bibitem{Fernandes:2019rez}
P.~G.~S.~Fernandes, C.~A.~R.~Herdeiro, A.~M.~Pombo, E.~Radu and N.~Sanchis-Gual,
Class. Quant. Grav. \textbf{36}, no.13, 134002 (2019)
[erratum: Class. Quant. Grav. \textbf{37}, no.4, 049501 (2020)]
doi:10.1088/1361-6382/ab23a1
[arXiv:1902.05079 [gr-qc]].

\bibitem{Santos:2020pmh}
N.~M.~Santos, C.~L.~Benone, L.~C.~B.~Crispino, C.~A.~R.~Herdeiro and E.~Radu,
JHEP \textbf{07}, 010 (2020)
doi:10.1007/JHEP07(2020)010
[arXiv:2004.09536 [gr-qc]].

\bibitem{MaAYu15}
A.K.M. Masood-ul-Alam and W. Yu, Communications in Analysis and Geometry, {\bf 23}, 377-387 (2015). 

\bibitem{Borghini:2019msu}
S.~Borghini, P.~T.~Chru\'sciel and L.~Mazzieri,
Arch. Ration. Mech. Anal. \textbf{247}, no.2, 1-35 (2023)
doi:10.1007/s00205-023-01860-1
[arXiv:1909.05941 [math.DG]].

\bibitem{Cai:1997ij}
R.~G.~Cai, J.~Y.~Ji, 
Phys. Rev. D \textbf{58}, 024002 (1998)
doi:10.1103/PhysRevD.58.024002
[arXiv:gr-qc/9708064 [gr-qc]].

\bibitem{An:2023bpb}
Y.~P.~An and L.~Li,
Eur. Phys. J. C \textbf{83}, no.7, 569 (2023)
doi:10.1140/epjc/s10052-023-11758-7
[arXiv:2301.06312 [gr-qc]].
 

\bibitem{Emparan:2008eg}
R.~Emparan and H.~S.~Reall,
Living Rev. Rel. \textbf{11}, 6 (2008)
doi:10.12942/lrr-2008-6
[arXiv:0801.3471 [hep-th]].

\bibitem{Hollands:2012xy}
S.~Hollands and A.~Ishibashi,
Class. Quant. Grav. \textbf{29}, 163001 (2012)
doi:10.1088/0264-9381/29/16/163001
[arXiv:1206.1164 [gr-qc]].

\bibitem{Ida:2011jm}
D.~Ida, A.~Ishibashi and T.~Shiromizu,
Prog. Theor. Phys. Suppl. \textbf{189}, 52-92 (2011)
doi:10.1143/PTPS.189.52
[arXiv:1105.3491 [hep-th]].

\bibitem{Maldacena:1997re}
J.~M.~Maldacena,
``The Large N limit of superconformal field theories and supergravity,''
Adv. Theor. Math. Phys. \textbf{2}, 231-252 (1998)
[arXiv:hep-th/9711200 [hep-th]].
%
\bibitem{Gubser:1998bc}
S.~S.~Gubser, I.~R.~Klebanov and A.~M.~Polyakov,
``Gauge theory correlators from noncritical string theory,''
Phys. Lett. B \textbf{428}, 105-114 (1998)
[arXiv:hep-th/9802109 [hep-th]].
%
\bibitem{Witten:1998qj}
E.~Witten,
``Anti-de Sitter space and holography,''
Adv. Theor. Math. Phys. \textbf{2}, 253-291 (1998)
[arXiv:hep-th/9802150 [hep-th]].

\bibitem{Torii:2001pg}
T.~Torii, K.~Maeda and M.~Narita,
Phys. Rev. D \textbf{64}, 044007 (2001)
doi:10.1103/PhysRevD.64.044007

\bibitem{Winstanley:1998sn}
E.~Winstanley,
Class. Quant. Grav. \textbf{16}, 1963-1978 (1999)
doi:10.1088/0264-9381/16/6/325
[arXiv:gr-qc/9812064 [gr-qc]].

\bibitem{Winstanley:2002jt}
E.~Winstanley,
Found. Phys. \textbf{33}, 111-143 (2003)
doi:10.1023/A:1022871809835
[arXiv:gr-qc/0205092 [gr-qc]].

\bibitem{Brihaye:2019gla}
Y.~Brihaye, C.~Herdeiro and E.~Radu,
Phys. Lett. B \textbf{802}, 135269 (2020)
doi:10.1016/j.physletb.2020.135269
[arXiv:1910.05286 [gr-qc]].

\bibitem{Sudarsky:2002mk}
D.~Sudarsky and J.~A.~Gonzalez,
Phys. Rev. D \textbf{67}, 024038 (2003)
doi:10.1103/PhysRevD.67.024038
[arXiv:gr-qc/0207069 [gr-qc]].

\bibitem{Gibbons:2002av}
G.~W.~Gibbons, D.~Ida and T.~Shiromizu,
Phys. Rev. Lett. \textbf{89}, 041101 (2002)
doi:10.1103/PhysRevLett.89.041101
[arXiv:hep-th/0206049 [hep-th]].

\bibitem{Anabalon:2012ta}
A.~Anabalon,
JHEP \textbf{06}, 127 (2012)
doi:10.1007/JHEP06(2012)127
[arXiv:1204.2720 [hep-th]].


\bibitem{Horowitz:2022hlz}
G.~T.~Horowitz, D.~Wang and X.~Ye,
Class. Quant. Grav. \textbf{39}, no.22, 225014 (2022)
doi:10.1088/1361-6382/ac994b
[arXiv:2206.08944 [hep-th]].

\bibitem{Maeda:2006pm}
H.~Maeda,
Phys. Rev. D \textbf{73}, 104004 (2006)
doi:10.1103/PhysRevD.73.104004
[arXiv:gr-qc/0602109 [gr-qc]].

\bibitem{Szabados:2009eka}
L.~B.~Szabados,
Living Rev. Rel. \textbf{12}, 4 (2009)
doi:10.12942/lrr-2009-4

\bibitem{Liu:2003bx}
C.~C.~M.~Liu and S.~T.~Yau,
Phys. Rev. Lett. \textbf{90}, 231102 (2003)
doi:10.1103/PhysRevLett.90.231102
[arXiv:gr-qc/0303019 [gr-qc]].
%
\bibitem{Liu:2004dc}
C.~C.~M.~Liu and S.~T.~Yau,
[arXiv:math/0412292 [math.DG]].

\bibitem{Ghosh:2023kge}
R.~Ghosh, S.~Sk and S.~Sarkar,
Phys. Rev. D \textbf{108}, no.4, L041501 (2023)
doi:10.1103/PhysRevD.108.L041501
[arXiv:2306.14193 [gr-qc]].

\bibitem{Johannsen:2011dh}
T.~Johannsen and D.~Psaltis,
Phys. Rev. D \textbf{83}, 124015 (2011)
doi:10.1103/PhysRevD.83.124015
[arXiv:1105.3191 [gr-qc]].

\bibitem{Hayward:2005gi}
S.~A.~Hayward,
Phys. Rev. Lett. \textbf{96}, 031103 (2006)
doi:10.1103/PhysRevLett.96.031103
[arXiv:gr-qc/0506126 [gr-qc]].

\bibitem{Culetu:2013fsa}
H.~Culetu,
[arXiv:1305.5964 [gr-qc]].

\bibitem{Culetu:2014lca}
H.~Culetu,
Int. J. Theor. Phys. \textbf{54}, no.8, 2855-2863 (2015)
doi:10.1007/s10773-015-2521-6
[arXiv:1408.3334 [gr-qc]].

\bibitem{Simpson:2019mud}
A.~Simpson and M.~Visser,
Universe \textbf{6}, no.1, 8 (2019)
doi:10.3390/universe6010008
[arXiv:1911.01020 [gr-qc]].

\bibitem{Xiang:2013sza}
L.~Xiang, Y.~Ling and Y.~G.~Shen,
Int. J. Mod. Phys. D \textbf{22}, 1342016 (2013)
doi:10.1142/S0218271813420169
[arXiv:1305.3851 [gr-qc]].

\bibitem{Dymnikova:1992ux}
I.~Dymnikova,
Gen. Rel. Grav. \textbf{24}, 235-242 (1992)
doi:10.1007/BF00760226

\bibitem{Kiselev:2002dx}
V.~V.~Kiselev,
Class. Quant. Grav. \textbf{20}, 1187-1198 (2003)
doi:10.1088/0264-9381/20/6/310
[arXiv:gr-qc/0210040 [gr-qc]].

\bibitem{Hod:2011aa}
S.~Hod,
Phys. Rev. D \textbf{84}, 124030 (2011)
doi:10.1103/PhysRevD.84.124030
[arXiv:1112.3286 [gr-qc]].

\bibitem{Martinez:2002ru}
C.~Martinez, R.~Troncoso and J.~Zanelli,
Phys. Rev. D \textbf{67}, 024008 (2003)
doi:10.1103/PhysRevD.67.024008
[arXiv:hep-th/0205319 [hep-th]].

\bibitem{Fernando:2016ksb}
S.~Fernando,
Int. J. Mod. Phys. D \textbf{26}, no.07, 1750071 (2017)
doi:10.1142/S0218271817500717
[arXiv:1611.05337 [gr-qc]].


\bibitem{Boucher:1983cv}
W.~Boucher, G.~W.~Gibbons and G.~T.~Horowitz,
Phys. Rev. D \textbf{30}, 2447 (1984)
doi:10.1103/PhysRevD.30.2447

\bibitem{Anderson:2002xb}
M.~T.~Anderson, P.~T.~Chrusciel and E.~Delay,
JHEP \textbf{10}, 063 (2002)
doi:10.1088/1126-6708/2002/10/063
[arXiv:gr-qc/0211006 [gr-qc]].

\bibitem{Hartnoll:2008vx}
S.~A.~Hartnoll, C.~P.~Herzog and G.~T.~Horowitz,
Phys. Rev. Lett. \textbf{101}, 031601 (2008)
doi:10.1103/PhysRevLett.101.031601
[arXiv:0803.3295 [hep-th]].





\end{thebibliography}
\end{document}